\documentclass[floatfix,aps,prl,twocolumn]{revtex4-2}

\usepackage[T1]{fontenc}
\usepackage{lipsum}  
\usepackage{graphicx}
\usepackage[english]{babel}
\usepackage{upgreek}
\usepackage{amsmath}
\usepackage{amssymb}
\usepackage{tabls}
\usepackage{dsfont}
\usepackage[colorlinks=true,citecolor=blue,linkcolor=red,urlcolor=blue]{hyperref}
\usepackage{physics}

\makeatletter
\def\bbl@set@language#1{%
	\edef\languagename{%
		\ifnum\escapechar=\expandafter`\string#1\@empty
		\else\string#1\@empty\fi}%
	\@ifundefined{babel@language@alias@\languagename}{}{%
		\edef\languagename{\@nameuse{babel@language@alias@\languagename}}%
	}%
	\select@language{\languagename}%
	\expandafter\ifx\csname date\languagename\endcsname\relax\else
	\if@filesw
	\protected@write\@auxout{}{\string\select@language{\languagename}}%
	\bbl@for\bbl@tempa\BabelContentsFiles{%
		\addtocontents{\bbl@tempa}{\xstring\select@language{\languagename}}}%
	\bbl@usehooks{write}{}%
	\fi
	\fi}
\newcommand{\DeclareLanguageAlias}[2]{%
	\global\@namedef{babel@language@alias@#1}{#2}%
}
\makeatother

\newcommand\varpm{\mathbin{\vcenter{\hbox{%
  \oalign{\hfil$\scriptstyle+$\hfil\cr
          \noalign{\kern-.3ex}
          $\scriptscriptstyle({-})$\cr}%
}}}}
\newcommand\varmp{\mathbin{\vcenter{\hbox{%
  \oalign{$\scriptstyle({+})$\cr
          \noalign{\kern-.3ex}
          \hfil$\scriptscriptstyle-$\hfil\cr}%
}}}}

\DeclareLanguageAlias{en}{english}

\begin{document}

\title{High-fidelity CNOT gate for spin qubits with asymmetric driving using virtual gates}

\author{Irina Heinz}
\email{irina.heinz@uni-konstanz.de}
\affiliation{Department of Physics, University of Konstanz, D-78457 Konstanz, Germany}
\author{Guido Burkard}
\email{guido.burkard@uni-konstanz.de}
\affiliation{Department of Physics, University of Konstanz, D-78457 Konstanz, Germany}


\begin{abstract}
    Recent experiments have demonstrated two-qubit fidelities above 99\%, however, theoretically, the fidelity of CNOT operations is limited by off-resonant driving described by off-diagonal terms in the system Hamiltonian. Here we investigate these off-diagonal contributions and we propose a fidelity improvement of several orders of magnitude by using asymmetric driving. Therefore, we provide a description of an ac virtual gate based on a simple capacitance model which not only enables a high fidelity CNOT but also crosstalk reduction when scaling up spin qubit devices to larger arrays.
\end{abstract}


\maketitle

\textit{Introduction.}
For the realization of quantum computing spin qubits~\cite{Loss_1998} in silicon quantum dots \cite{Zwanenburg_2013} have become of great interest. High purification of silicon enables long qubit coherence times due to the weak spin-orbit interaction and so represents an interesting candidate for scaling up spin qubit arrays. A magnetic gradient field induced by a micromagnet \cite{Yoneda_2015, Kawakami_2014} allows for separate splitting energies and thus individual addressability of the spins. Using electric dipole spin resonance (EDSR) by modulating the quantum dot defining gate voltages, leading to a modulation of the confined electron within the magnetic gradient field and thus an effective magnetic drive, single qubit gates can be performed. Two-qubit gates are realized by switching on the exchange interaction between neighboring qubits \cite{PhysRevB.59.2070, Yoneda_2017, Watson_2018, Zajac_2017, PhysRevLett.107.146801}. If operating at a symmetric operation point first order charge noise can be suppressed~\cite{PhysRevLett.116.116801, PhysRevLett.116.110402, PhysRevLett.115.096801}.

While high-fidelity single-qubit operations have already been demonstrated, recent results show fidelities of $>$99.5\% also for two-qubit gates \cite{noiri2021fast, xue2021computing}. For the CNOT gate implementation in Ref. \cite{noiri2021fast} the exchange interaction between two spins combined with an oscillating magnetic drive on both qubits results in a high-fidelity operation when synchronizing the off-resonant Rabi oscillation of the nearby transition \cite{Russ_2018,Zajac_2017}. However, in this previous description the CNOT operation suffers from an upper theoretic fidelity bound due to neglected off-diagonal parts of the Hamiltonian. Here, we analyze these off-diagonal elements and find an asymmetric driving within the CNOT to reduce the infidelity of several magnitudes from $10^{-3}$ to $10^{-6}$. To realize such a high-fidelity CNOT implementation we provide a description for an ac driven virtual gate based on a simple capacitance model.


\textit{Theoretical Model.}
First, we consider a gate defined double quantum dot (DQD) in the $(1,1)$ charge regime, where we neglect excited valley states and intrinsic spin-orbit coupling. The system can be described theoretically by the Heisenberg Hamiltonian $H = J (t) \left( \mathbf{S}^L \cdot \mathbf{S}^R -1/4 \right) + \mathbf{S}^L \cdot \mathbf{B}^L + \mathbf{S}^R \cdot \mathbf{B}^R$, where $J$ is the tunable exchange interaction between spins $\mathbf{S}^{L}$ and $\mathbf{S}^{R}$ tuned by the middle barrier gate, such as in Refs.~\cite{Russ_2018} and \cite{Zajac_2017}, required for two-qubit operations, and $\mathbf{B}_{\alpha} = (0,B_{y}^{\alpha}(t),B_{z}^{\alpha})$ is the external magnetic field at the position of spin $\mathbf{S}^{\alpha}$, where $\alpha \in \{L,R\}$. Magnetic fields are represented in energy units throughout this paper, i.e. $\mathbf{B}_{\mathrm{physical}}=\mathbf{B}/g\mu_B$, and we furthermore set $\hbar = 1$. 
A large homogeneous magnetic field and a field gradient in $z$-direction along the $x$-axis, e.g., caused by a micromagnet, $B_{z}^{\alpha}=B_{z} + b_{z}^{\alpha}$, allows individual addressability of single spins, and a small field gradient in $y$-direction enables a time dependent EDSR driving field in $y$-direction $B_{y}^{\alpha}(t) = B_{y,0}^{\alpha} + B_{y,1}^{\alpha} \cos(\omega t + \theta)$ when oscillating plunger gate voltages. The amplitude of the EDSR induced effective magnetic driving strength for EDSR is proportional to the electric field and depends on the device architecture, natural or artificial spin-orbit coupling mechanism, and applied gate voltage \cite{PhysRevB.74.165319, PhysRevLett.96.047202}.
We define $E_z = (B_z^L + B_z^R)/2$ and $\Delta E_z = B_z^R - B_z^L$ for the remainder of this paper, so in the regime of weak exchange ($J\ll\delta E_z$) the slightly corrected states $\{\ket{\uparrow \uparrow}, \tilde{\ket{\downarrow \uparrow}}, \tilde{\ket{\uparrow \downarrow}}, \ket{\downarrow \downarrow}\}$ are the eigenstates of the Hamiltonian with instantaneous eigenvalues $\mathcal{E} (\ket{\uparrow \uparrow}) = E_z$, 
$\mathcal{E} (\tilde{\ket{\uparrow \downarrow}}) = \frac{1}{2} (-J-\sqrt{J^2+\Delta E_z^2})$,
$\mathcal{E} (\tilde{\ket{\downarrow \uparrow}}) = \frac{1}{2} (-J+\sqrt{J^2+\Delta E_z^2})$, 
$\mathcal{E} (\ket{\downarrow \downarrow}) = -E_z$ \cite{Russ_2018}.
This shift of energy levels allows individual addressing of the $\ket{\uparrow \uparrow} \leftrightarrow \tilde{\ket{\downarrow \uparrow}}$ transition at the resonance frequency $\omega_{\text{CNOT}} = E_z + (J - \sqrt{\Delta E_z^2+J^2})/2$ due to the distinctness of the available transition frequencies. 
In the rotating frame $\tilde{H}(t) = R^{\dagger}HR+i\dot{R}^{\dagger}R$ with $R=\exp(-i \omega t (\mathbf{S}^L + \mathbf{S}^R))$ we make the rotating wave approximation (RWA) in which far off-resonant oscillations can be neglected. When further approximating $\sqrt{J^2+\Delta E_z^2} \approx \Delta E_z + J^2/\Delta E_z$ and $(J/\Delta E_z)^2 \approx 0$, since $J\ll\Delta E_z$ holds, we can bring the Hamiltonian in the instantaneous eigenbasis into the form
\begin{align}
    \tilde{H} \approx \frac{1}{2}
    \left(\begin{array}{cc}
    A_+ & B^{\dagger}\\
    B & A_-
    \end{array}\right) \label{Hamilton2}
\end{align}
with 
\begin{align}
    A_{\pm} =
    \left(\begin{array}{cc}
    \pm 2 (E_z \mp \omega) & \mp i \alpha_{\pm}^*\\
    \pm i \alpha_{\pm} & -J \pm \left(\Delta E_z + \frac{J^2}{2\Delta E_z}\right)
    \end{array}\right) \label{Hamilton2}
\end{align}
representing the resonant and off-resonant oscillating parts and coupling matrix
\begin{align}
    B =
    \left(\begin{array}{cc}
    0 & i \beta_+\\
    i \beta_- & 0
    \end{array}\right) \label{Hamilton2}
\end{align}
where
\begin{align}
    \alpha_{\pm} = \left(\pm B_{y,1}^L + B_{y,1}^R \frac{J}{2\Delta E_z} \right) e^{\pm i\theta},\\
    \beta_{\pm} = \left(\pm B_{y,1}^R + B_{y,1}^L \frac{J}{2\Delta E_z} \right) e^{i\theta}. \label{beta}
\end{align}

In Ref.~\cite{Russ_2018} high-fidelity CNOT implementations were proposed by neglecting $\beta_+$ and $\beta_-$ to decouple the block matrices into the upper left and the lower right part of the matrix. By synchronizing the resonant Rabi frequency $\Omega = |\alpha_+|$ and off-resonant frequency $\tilde{\Omega} = \sqrt{|\alpha_-|^2+J^2}$ such that the off-resonant block matrix evolves full $2\pi$ rotations yields $\Omega = \frac{2m+1}{2n} \tilde{\Omega}\hspace*{0.5cm} m, n\in\mathbb{Z}$. This was also demonstrated experimentally in Ref. \cite{Zajac_2017} and \cite{noiri2021fast}. Different from this previous work we allow driving strengths on the left and right qubit to be different, and thus, the synchronization yields
\begin{widetext}
    \begin{align}
        B_{y,1}^L = \frac{J}{2\Delta E_z} \frac{4(2m+1)\sqrt{(4n^2 - (2m+1)^2)(\Delta E_z)^2 + n^2(B_{y,1}^R)^2} - \left(4n^2 + (2m+1)^2\right)B_{y,1}^R}{4n^2 - (2m+1)^2} .\label{eq:BLofBRandJ}
    \end{align}
\end{widetext}
The driving field on the left qubit in Equation~\eqref{eq:BLofBRandJ} depends not only on $J$ but also on the choice of $B_{y,1}^R$ and represents the main result of this paper. This leaves an additional parameter to fine-tune the CNOT quality, which can be evaluated by calculating the fidelity~\cite{Pedersen_2007} $F~=~(d+| \text{Tr}[ U_{\text{ideal}}^{\dagger}U_{\text{actual}} ] |^2 )/(d(d+1))$,
where $d$ is the dimension of the Hilbert space, $U_{\text{ideal}}$ is the desired CNOT operation and $U_{\text{actual}}=\text{exp}(-i\tilde{H}t)$ the actual operation. The gate time $\tau_{\text{CNOT}} = \pi (2m+1)/|\alpha_+|$ is mainly determined by the large magnetic gradient $\Delta E_z$ but changes for different $B_{y,1}^{R}$.

\begin{figure}[t]
	\centering
	\includegraphics[width=0.47\textwidth]{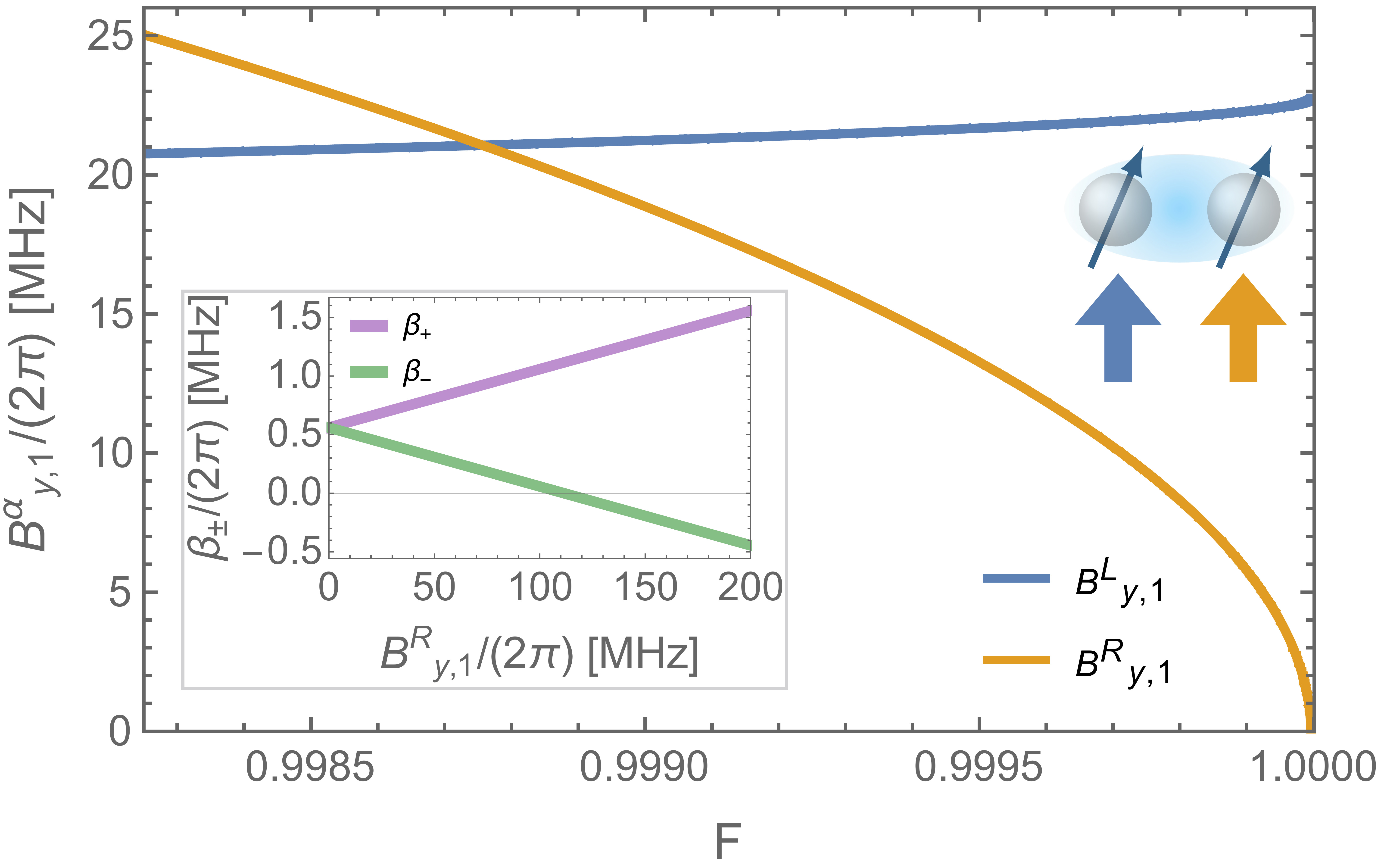}
	\caption{CNOT fidelity dependence on the driving strength $B_{y,1}^R$ (orange), which determines the choice of $B_{y,1}^L$ (blue). The inset shows values for $\beta_+$ (purple) and $\beta_-$ (light green) depending on $B_{y,1}^R$ where $\theta = 0$. $B_{y,1}^L$ is determined by Equation~\eqref{eq:BLofBRandJ}, i.e. by the choice of $B_{y,1}^R$ and $J = (2\pi)\,19.7$ MHz, where $n=1$ and $m=0$.}
	\label{fig:CNOTasym}
\end{figure}

\textit{Asymmetrically driven CNOT gate.}
Taking the CNOT synchronization condition of Ref.~\cite{Russ_2018} into account, which indeed maximizes the qubit gate fidelity, we take a look at the absolute value of the coefficients $\beta_+$ and $\beta_-$ (set $\theta = 0$), which represent the fidelity limiting factors in the noiseless CNOT gate operation. Considering Equation~\eqref{beta} for independently chosen $B_{y,1}^L$ and $B_{y,1}^R$, we find that $\beta_+$ and $\beta_-$ are equal, i.e. $\beta_+=\beta_-=B_{y,1}^L J/(2\Delta E_z)$, if $B_{y,1}^R=0$, and we obtain $\beta_{\pm}=0$ and $\beta_{\mp}=B_{y,1}^L J/\Delta E_z$ for $B_{y,1}^R = \mp B_{y,1}^L J/(2\Delta E_z)$. Regarding the conditional choice of the left qubit's driving strength $B_{y,1}^L$ depending on the right qubit's driving $B_{y,1}^R$ we calculate the values for $\beta_+$ and $\beta_-$ in the inset of Fig. \ref{fig:CNOTasym}. Here we used $n=1$, $m=0$ and $J=(2\pi)\,19.7$~MHz as in the experiment in Ref. \cite{Zajac_2017}. In order to keep the off-diagonal elements small we restrict the choice of the right driving strength to $< (2\pi)\,120$~MHz. 

Using these values and $\theta=3\pi/2$ the CNOT gate fidelity is calculated and shown in Fig. \ref{fig:CNOTasym}. Although for $B_{y,1}^R \approx (2\pi)\,111$~MHz one of the off-diagonal elements becomes zero, we find that the fidelity decreases since the remaining element $\beta_+$ becomes twice as large as in case of no driving on the right dot. Obviously, small fields $B_{y,1}^R$ lead to higher fidelity and thus are shown in Fig.~\ref{fig:CNOTasym}. The choice of zero driving on the control qubit can even minimize the infidelity down to $10^{-6}$, where instead of the driving time of $284$~ns for symmetric driving, we obtain a time of $276$~ns for asymmetric driving. The remaining infidelity is due to finite $|\beta_+|=|\beta_-|=B_{y,1}^L J/(2\Delta E_z)$ and in principle can be reduced by a high magnetic gradient field or a smaller value for $J$. This corresponds to only driving the target qubit while leaving the control qubit alone.
\begin{figure}[ht]
	\centering
	\includegraphics[width=0.47\textwidth]{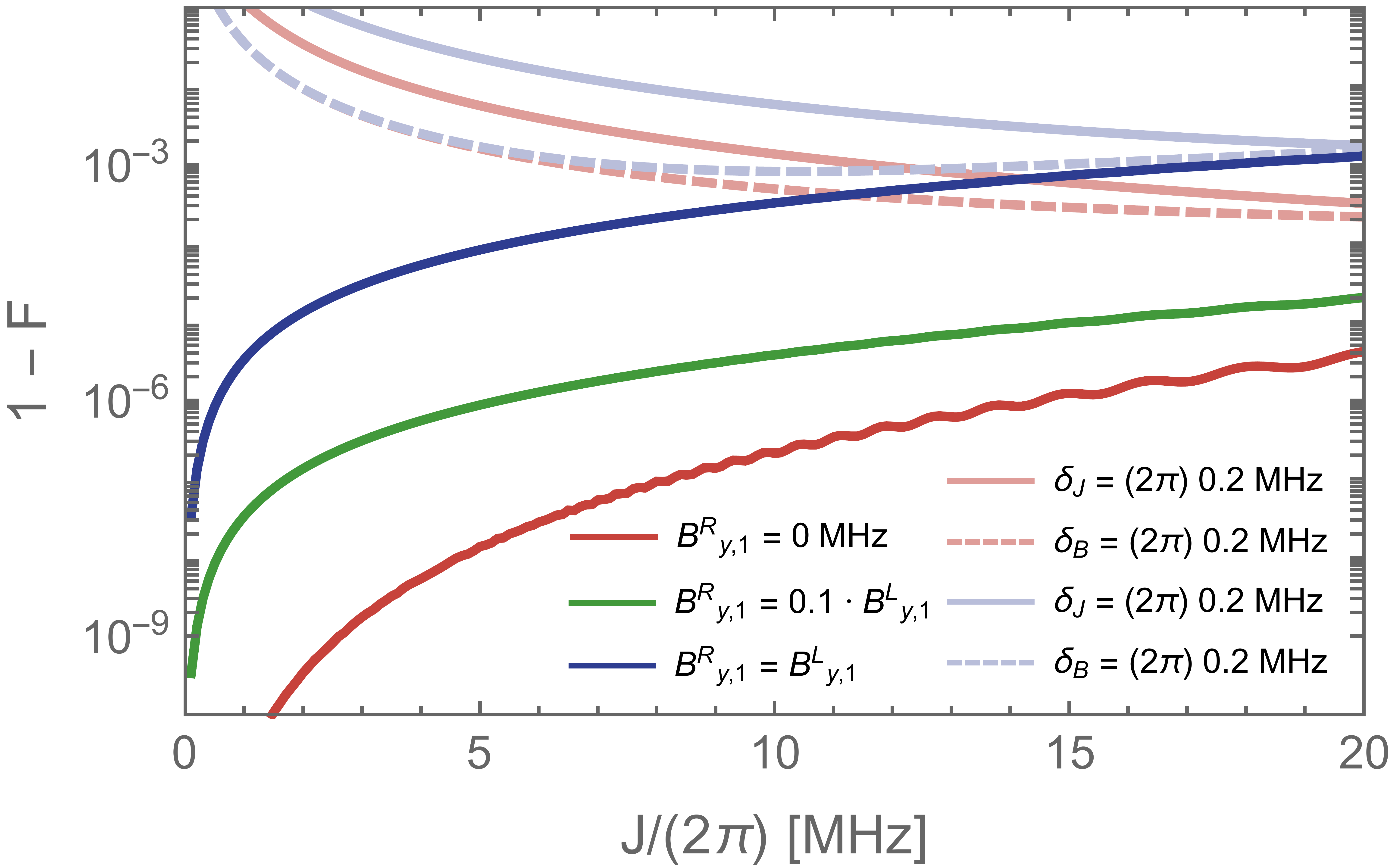}
	\caption{Infidelity depending on $J$ for symmetric driving ($B_{y,1}^R = B_{y,1}^L$) in dark blue, and asymmetric driving $B_{y,1}^R = 0$~MHz (dark red) and $B_{y,1}^R = 0.1 \cdot B_{y,1}^L$ (green) with $n=1$ and $m=0$. The infidelity of symmetrically and asymmetrically ($B_{y,1}^R = 0$~MHz) driven CNOT gates suffering from Gaussian charge and dephasing noise with standard deviations $\delta_J = (2\pi)\,0.2$~MHz and $\delta_B = (2\pi)\,0.2$~MHz are shown in light blue and light red solid and dashed lines, respectively.}
	\label{fig:I-J}
\end{figure}
Fig.~\ref{fig:I-J} shows the infidelity of the CNOT operation depending on $J$ for the case of symmetric driving $B_{y,1}^R = B_{y,1}^L$ in dark blue and asymmetric driving with $B_{y,1}^R = 0$~MHz in dark red and $B_{y,1}^R = 0.1 \cdot B_{y,1}^L$ in green. In each case the infidelity increases for larger $J$, but in the asymmetric cases this behaviour is strongly suppressed and becomes only noticeable as infidelities with magnitudes of $10^{-6}$ and $10^{-5}$. Additionally, an oscillating behaviour with small amplitude shows up in each of the plots. These are due to interpolations of the off-resonant Rabi oscillations resulting from finite $|\beta_+|$ and $|\beta_-|$.
We also consider a Gaussian noise acting on the exchange interaction $J$ in terms of charge noise with standard deviation $\delta_J = (2\pi)\,0.2$~MHz. In Fig.~\ref{fig:I-J} the noisy symmetrically driven CNOT gate is shown as a light blue solid line and the asymmetric case with $B_{y,1}^R = 0$~MHz as a light red solid line. For small values for $J$ the noise contribution is large compared to the exchange interaction and thus behaves similar in both cases, however, for larger exchange couplings both curves approach their noise-free infidelity curves. Thus, we find the noisy asymmetrically driven CNOT gate to perform even better than the noise-free symmetrically driven CNOT gate. We find, that the same holds for dephasing noise, which is taken into account in Fig.~\ref{fig:I-J} as light blue and light red dashed lines for the symmetric and asymmetric case, respectively. We again assume a Gaussian noise with standard deviation $\delta_B = (2\pi)\,0.2$~MHz, where we consider independent fluctuations of $B_{z,1}$ and $B_{z,2}$. This corresponds to dephasing times similar as in Ref.~\cite{noiri2021fast}. Similar to the case of charge noise the infidelity decreases first and then approaches the noise-free line, such that the noisy asymmetrically driven CNOT operation performs better than the noise-free symmetric CNOT. For $J=(2\pi)\,20$~MHz we find the difference to be of at least one magnitude. When considering a system with even less noise and thus smaller standard deviations $\delta_J$ and $\delta_B$ the noisy asymmetric CNOT gate performs even better and is only limited by the theoretical bound due to the off-diagonal elements, which in our case is of magnitude $10^{-6}$.

\begin{figure}[t]
	\centering
	\includegraphics[width=0.35\textwidth]{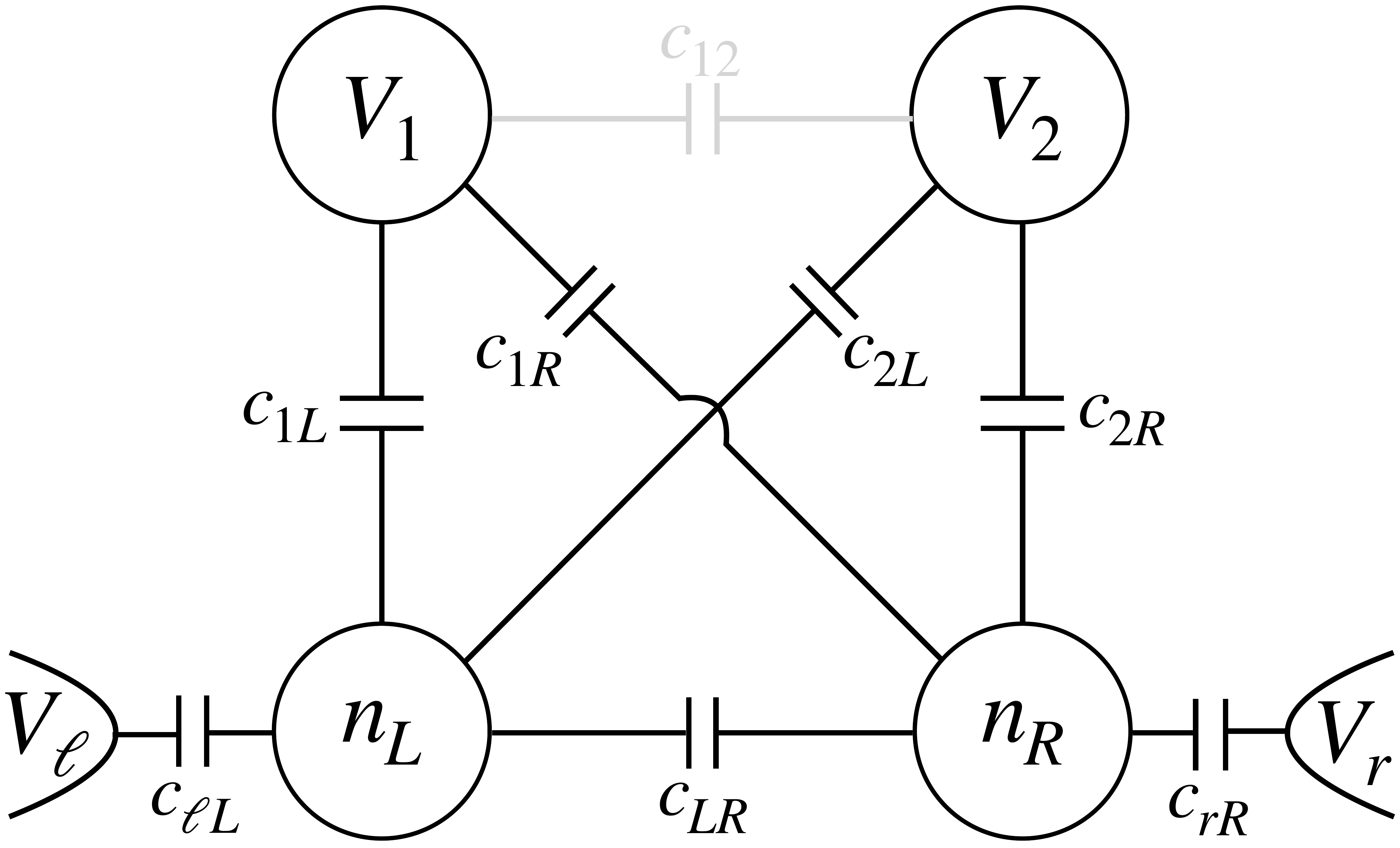}
	\caption{Capacitance model to realize an ac driven virtual gate: The average electron occupation number of the left and right dots $\langle n_L\rangle = 1$ and $\langle n_R\rangle = 1$ are constant while the induced charge on the capacitors are time dependent.}
	\label{fig:CapacitanceModel}
\end{figure}

\textit{Virtual ac gates.}
In order to realize asymmetric driving of a CNOT gate we introduce a linear capacitance model in Fig. \ref{fig:CapacitanceModel} to implement virtual gates for ac driving. The left and right quantum dots are labeled by their electron occupation number $n_L$ and $n_R$ and plunger gates by their gate voltages $V_1$ and $V_2$. Each quantum dot is coupled to each gate, to the neighboring dot and to a lead on the right ($V_r$) for the right dot and on the left ($V_\ell$) for the left dot, respectively. 
Using this simplified macroscopic model a static voltage offset determines the occupation number of one electron per dot and the time-dependent part enables the EDSR drive to manipulate the spin.  The charge on the left and right dots and thus the sum of charges induced by the conductive coupling is constant $\sum_{\gamma} \dot{Q}_{\gamma} =0$ ($\gamma \in \{\ell L, 1L, 2L, LR\}$ and $\{rR, 1R, 2R, RL\}$ and $\dot{Q}_{RL} = - \dot{Q}_{LR}$). In the slow relaxation regime (i.e. spin and charge relaxation rates $\Gamma_{1}^c,\Gamma_{1}^s \ll \omega$) \cite{PhysRevB.95.045414} and with $Q_{\gamma}=c_{\gamma} V_{\gamma}$ this leads to 
\begin{align}
    c_{\alpha} \dot{V}_{\alpha} = c_{1 \alpha}\dot{V}_1 + c_{2 \alpha} \dot{V}_2 + c_{LR} \dot{V}_{\overline{\alpha}} ,
\end{align}
where $\alpha = L,R$ and $\overline{\alpha}= R,L$ and $c_L= c_{\ell L} + c_{1 L} + c_{2 L} + c_{LR}$ and $c_R= c_{rR} + c_{1 R} + c_{2 R} + c_{LR}$. $V_L$ and $V_R$ are the voltages at the left and right dot respectively. We assume $V_{\ell}$ and $V_r$ to be constant and obtain the derivative solutions 
\begin{align}
    \dot{V}_L = \frac{\left( c_{1L} c_{R} + c_{1R} c_{LR} \right) \dot{V}_1 + \left( c_{2L} c_{R} + c_{2R} c_{LR} \right) \dot{V}_2}{c_L c_R - c_{LR}^2},\\
    \dot{V}_R = \frac{\left( c_{2R} c_{L} + c_{2L} c_{LR} \right) \dot{V}_2 + \left( c_{1R} c_{L} + c_{1L} c_{LR} \right) \dot{V}_1}{c_L c_R - c_{LR}^2}.
\end{align}
Hence, the solution of $V_L$ and $V_R$ linearly depends on the driving strengths $V_1$ and $V_2$ plus a constant part. To obtain an asymmetric CNOT drive we set $V_R = 0$ leading to a dependent choice of the two driving strengths
\begin{align}
    \dot{V}_2 = -\frac{c_{1R} c_{L} + c_{1L} c_{LR}}{c_{2R} c_{L} + c_{2L} c_{LR}} \dot{V}_1 ,
\end{align}
and thus an actual driving of the left dot of 
\begin{align}
    \dot{V}_L = \frac{c_{1L} c_{2R} - c_{1R} c_{2L}}{c_{2R} c_{L} + c_{2L} c_{LR}} \dot{V}_1 .
\end{align}
Applying a cosine drive leads to a simple linear dependence between dot voltage and plunger gate voltages and can simply be realised after measuring the respective capacitances.
Analogous to the asymmetric CNOT in a DQD, this way ac driven virtual gates can also be implemented on larger arrays to reduce crosstalk effects~\cite{PhysRevB.104.045420, heinz2021crosstalk}.

The influence of the inductance of gate electrodes on an ac driven signal can be estimated by the magnitudes of capacitances ($10^{-18}$ F) \cite{Russ_2020}, the length $l$ and radius $r$ of the gates and the distance $d$ between gates, which we assume to be of same magnitude within the $10^{-7}$ m regime. At frequencies lower than $\omega_0=1/\sqrt{LC}$ the capacitive impedance dominates the circuit while higher frequencies increase the role of inductive impedance. With $L\propto l$ and $L\propto ln((d-r)/r)$, we estimate $L\approx 10^{-13} H$ and thus $\omega_0 \approx 10^{15}$ Hz. Since we apply pulses with frequencies of several GHz, we here neglect the inductive impedance.

\textit{Conclusions}
In this paper we have shown a high-fidelity CNOT gate implementation using an asymmetric ac drive, where we operate only the target qubit while leaving the control qubit alone. This way off-diagonal terms in the Hamiltonian leading to off-resonant Rabi oscillations are minimized and enable higher fidelities for the CNOT gate. With our implementation the infidelity decreases of several magnitudes to $10^{-6}$ in the noise-free case. For noisy spin qubits we predict a fidelity improvement of at least about one magnitude compared to the symmetrically driven CNOT gate, which suggests a higher fidelity in experimental setups than so far demonstrated

We have further presented an estimation for the description of virtual gates using ac drives for frequencies below the critical frequency $\omega_0$ determined by inductances and capacitances within the gate defined quantum dot architecture and thus enable the realization of the asymmetrically driven CNOT gate. Additionally, this implementation of ac virtual gates can be generalized to the application to larger arrays which allows a scheme for the cancellation of crosstalk effects \cite{PhysRevB.104.045420, heinz2021crosstalk} and thus high fidelity quantum operations within large qubit arrays, and so becomes an interesting method for scaling up spin qubit devices to realize quantum computation.

\textit{Acknowledgments.}
This work has been supported by QLSI with funding from the European Union's Horizon 2020 research and innovation programme under grant agreement No 951852 and by the Deutsche Forschungsgemeinschaft (DFG, German Research Foundation) Grant No. SFB 1432 - Project-ID 425217212.

\bibliography{bibliography}

\begin{thebibliography}{22}%
\makeatletter
\providecommand \@ifxundefined [1]{%
 \@ifx{#1\undefined}
}%
\providecommand \@ifnum [1]{%
 \ifnum #1\expandafter \@firstoftwo
 \else \expandafter \@secondoftwo
 \fi
}%
\providecommand \@ifx [1]{%
 \ifx #1\expandafter \@firstoftwo
 \else \expandafter \@secondoftwo
 \fi
}%
\providecommand \natexlab [1]{#1}%
\providecommand \enquote  [1]{``#1''}%
\providecommand \bibnamefont  [1]{#1}%
\providecommand \bibfnamefont [1]{#1}%
\providecommand \citenamefont [1]{#1}%
\providecommand \href@noop [0]{\@secondoftwo}%
\providecommand \href [0]{\begingroup \@sanitize@url \@href}%
\providecommand \@href[1]{\@@startlink{#1}\@@href}%
\providecommand \@@href[1]{\endgroup#1\@@endlink}%
\providecommand \@sanitize@url [0]{\catcode `\\12\catcode `\$12\catcode
  `\&12\catcode `\#12\catcode `\^12\catcode `\_12\catcode `\%12\relax}%
\providecommand \@@startlink[1]{}%
\providecommand \@@endlink[0]{}%
\providecommand \url  [0]{\begingroup\@sanitize@url \@url }%
\providecommand \@url [1]{\endgroup\@href {#1}{\urlprefix }}%
\providecommand \urlprefix  [0]{URL }%
\providecommand \Eprint [0]{\href }%
\providecommand \doibase [0]{https://doi.org/}%
\providecommand \selectlanguage [0]{\@gobble}%
\providecommand \bibinfo  [0]{\@secondoftwo}%
\providecommand \bibfield  [0]{\@secondoftwo}%
\providecommand \translation [1]{[#1]}%
\providecommand \BibitemOpen [0]{}%
\providecommand \bibitemStop [0]{}%
\providecommand \bibitemNoStop [0]{.\EOS\space}%
\providecommand \EOS [0]{\spacefactor3000\relax}%
\providecommand \BibitemShut  [1]{\csname bibitem#1\endcsname}%
\let\auto@bib@innerbib\@empty
\bibitem [{\citenamefont {Loss}\ and\ \citenamefont
  {DiVincenzo}(1998)}]{Loss_1998}%
  \BibitemOpen
  \bibfield  {author} {\bibinfo {author} {\bibfnamefont {D.}~\bibnamefont
  {Loss}}\ and\ \bibinfo {author} {\bibfnamefont {D.~P.}\ \bibnamefont
  {DiVincenzo}},\ }\bibfield  {title} {\bibinfo {title} {Quantum computation
  with quantum dots},\ }\href {https://doi.org/10.1103/physreva.57.120}
  {\bibfield  {journal} {\bibinfo  {journal} {Phys. Rev. A}\ }\textbf {\bibinfo
  {volume} {57}},\ \bibinfo {pages} {120} (\bibinfo {year} {1998})}\BibitemShut
  {NoStop}%
\bibitem [{\citenamefont {Zwanenburg}\ \emph {et~al.}(2013)\citenamefont
  {Zwanenburg}, \citenamefont {Dzurak}, \citenamefont {Morello}, \citenamefont
  {Simmons}, \citenamefont {Hollenberg}, \citenamefont {Klimeck}, \citenamefont
  {Rogge}, \citenamefont {Coppersmith},\ and\ \citenamefont
  {Eriksson}}]{Zwanenburg_2013}%
  \BibitemOpen
  \bibfield  {author} {\bibinfo {author} {\bibfnamefont {F.~A.}\ \bibnamefont
  {Zwanenburg}}, \bibinfo {author} {\bibfnamefont {A.~S.}\ \bibnamefont
  {Dzurak}}, \bibinfo {author} {\bibfnamefont {A.}~\bibnamefont {Morello}},
  \bibinfo {author} {\bibfnamefont {M.~Y.}\ \bibnamefont {Simmons}}, \bibinfo
  {author} {\bibfnamefont {L.~C.~L.}\ \bibnamefont {Hollenberg}}, \bibinfo
  {author} {\bibfnamefont {G.}~\bibnamefont {Klimeck}}, \bibinfo {author}
  {\bibfnamefont {S.}~\bibnamefont {Rogge}}, \bibinfo {author} {\bibfnamefont
  {S.~N.}\ \bibnamefont {Coppersmith}},\ and\ \bibinfo {author} {\bibfnamefont
  {M.~A.}\ \bibnamefont {Eriksson}},\ }\bibfield  {title} {\bibinfo {title}
  {Silicon quantum electronics},\ }\href
  {https://doi.org/10.1103/revmodphys.85.961} {\bibfield  {journal} {\bibinfo
  {journal} {Rev. Mod. Phys.}\ }\textbf {\bibinfo {volume} {85}},\ \bibinfo
  {pages} {961} (\bibinfo {year} {2013})}\BibitemShut {NoStop}%
\bibitem [{\citenamefont {Yoneda}\ \emph {et~al.}(2015)\citenamefont {Yoneda},
  \citenamefont {Otsuka}, \citenamefont {Takakura}, \citenamefont
  {Pioro-Ladrière}, \citenamefont {Brunner}, \citenamefont {Lu}, \citenamefont
  {Nakajima}, \citenamefont {Obata}, \citenamefont {Noiri}, \citenamefont
  {Palmstrøm},\ and\ \citenamefont {et~al.}}]{Yoneda_2015}%
  \BibitemOpen
  \bibfield  {author} {\bibinfo {author} {\bibfnamefont {J.}~\bibnamefont
  {Yoneda}}, \bibinfo {author} {\bibfnamefont {T.}~\bibnamefont {Otsuka}},
  \bibinfo {author} {\bibfnamefont {T.}~\bibnamefont {Takakura}}, \bibinfo
  {author} {\bibfnamefont {M.}~\bibnamefont {Pioro-Ladrière}}, \bibinfo
  {author} {\bibfnamefont {R.}~\bibnamefont {Brunner}}, \bibinfo {author}
  {\bibfnamefont {H.}~\bibnamefont {Lu}}, \bibinfo {author} {\bibfnamefont
  {T.}~\bibnamefont {Nakajima}}, \bibinfo {author} {\bibfnamefont
  {T.}~\bibnamefont {Obata}}, \bibinfo {author} {\bibfnamefont
  {A.}~\bibnamefont {Noiri}}, \bibinfo {author} {\bibfnamefont {C.~J.}\
  \bibnamefont {Palmstrøm}},\ and\ \bibinfo {author} {\bibnamefont {et~al.}},\
  }\bibfield  {title} {\bibinfo {title} {Robust micromagnet design for fast
  electrical manipulations of single spins in quantum dots},\ }\href
  {https://doi.org/10.7567/apex.8.084401} {\bibfield  {journal} {\bibinfo
  {journal} {Appl. Phys. Express}\ }\textbf {\bibinfo {volume} {8}},\ \bibinfo
  {pages} {084401} (\bibinfo {year} {2015})}\BibitemShut {NoStop}%
\bibitem [{\citenamefont {Kawakami}\ \emph {et~al.}(2014)\citenamefont
  {Kawakami}, \citenamefont {Scarlino}, \citenamefont {Ward}, \citenamefont
  {Braakman}, \citenamefont {Savage}, \citenamefont {Lagally}, \citenamefont
  {Friesen}, \citenamefont {Coppersmith}, \citenamefont {Eriksson},\ and\
  \citenamefont {Vandersypen}}]{Kawakami_2014}%
  \BibitemOpen
  \bibfield  {author} {\bibinfo {author} {\bibfnamefont {E.}~\bibnamefont
  {Kawakami}}, \bibinfo {author} {\bibfnamefont {P.}~\bibnamefont {Scarlino}},
  \bibinfo {author} {\bibfnamefont {D.~R.}\ \bibnamefont {Ward}}, \bibinfo
  {author} {\bibfnamefont {F.~R.}\ \bibnamefont {Braakman}}, \bibinfo {author}
  {\bibfnamefont {D.~E.}\ \bibnamefont {Savage}}, \bibinfo {author}
  {\bibfnamefont {M.~G.}\ \bibnamefont {Lagally}}, \bibinfo {author}
  {\bibfnamefont {M.}~\bibnamefont {Friesen}}, \bibinfo {author} {\bibfnamefont
  {S.~N.}\ \bibnamefont {Coppersmith}}, \bibinfo {author} {\bibfnamefont
  {M.~A.}\ \bibnamefont {Eriksson}},\ and\ \bibinfo {author} {\bibfnamefont
  {L.~M.~K.}\ \bibnamefont {Vandersypen}},\ }\bibfield  {title} {\bibinfo
  {title} {Electrical control of a long-lived spin qubit in a {S}i/{S}i{G}e
  quantum dot},\ }\href {https://doi.org/10.1038/nnano.2014.153} {\bibfield
  {journal} {\bibinfo  {journal} {Nat. Nanotechnol.}\ }\textbf {\bibinfo
  {volume} {9}},\ \bibinfo {pages} {666} (\bibinfo {year} {2014})}\BibitemShut
  {NoStop}%
\bibitem [{\citenamefont {Burkard}\ \emph {et~al.}(1999)\citenamefont
  {Burkard}, \citenamefont {Loss},\ and\ \citenamefont
  {DiVincenzo}}]{PhysRevB.59.2070}%
  \BibitemOpen
  \bibfield  {author} {\bibinfo {author} {\bibfnamefont {G.}~\bibnamefont
  {Burkard}}, \bibinfo {author} {\bibfnamefont {D.}~\bibnamefont {Loss}},\ and\
  \bibinfo {author} {\bibfnamefont {D.~P.}\ \bibnamefont {DiVincenzo}},\
  }\bibfield  {title} {\bibinfo {title} {Coupled quantum dots as quantum
  gates},\ }\href {https://doi.org/10.1103/PhysRevB.59.2070} {\bibfield
  {journal} {\bibinfo  {journal} {Phys. Rev. B}\ }\textbf {\bibinfo {volume}
  {59}},\ \bibinfo {pages} {2070} (\bibinfo {year} {1999})}\BibitemShut
  {NoStop}%
\bibitem [{\citenamefont {Yoneda}\ \emph {et~al.}(2017)\citenamefont {Yoneda},
  \citenamefont {Takeda}, \citenamefont {Otsuka}, \citenamefont {Nakajima},
  \citenamefont {Delbecq}, \citenamefont {Allison}, \citenamefont {Honda},
  \citenamefont {Kodera}, \citenamefont {Oda}, \citenamefont {Hoshi} \emph
  {et~al.}}]{Yoneda_2017}%
  \BibitemOpen
  \bibfield  {author} {\bibinfo {author} {\bibfnamefont {J.}~\bibnamefont
  {Yoneda}}, \bibinfo {author} {\bibfnamefont {K.}~\bibnamefont {Takeda}},
  \bibinfo {author} {\bibfnamefont {T.}~\bibnamefont {Otsuka}}, \bibinfo
  {author} {\bibfnamefont {T.}~\bibnamefont {Nakajima}}, \bibinfo {author}
  {\bibfnamefont {M.~R.}\ \bibnamefont {Delbecq}}, \bibinfo {author}
  {\bibfnamefont {G.}~\bibnamefont {Allison}}, \bibinfo {author} {\bibfnamefont
  {T.}~\bibnamefont {Honda}}, \bibinfo {author} {\bibfnamefont
  {T.}~\bibnamefont {Kodera}}, \bibinfo {author} {\bibfnamefont
  {S.}~\bibnamefont {Oda}}, \bibinfo {author} {\bibfnamefont {Y.}~\bibnamefont
  {Hoshi}}, \emph {et~al.},\ }\bibfield  {title} {\bibinfo {title} {A
  quantum-dot spin qubit with coherence limited by charge noise and fidelity
  higher than 99.9\%},\ }\href {https://doi.org/10.1038/s41565-017-0014-x}
  {\bibfield  {journal} {\bibinfo  {journal} {Nat. Nanotechnol.}\ }\textbf
  {\bibinfo {volume} {13}},\ \bibinfo {pages} {102} (\bibinfo {year}
  {2017})}\BibitemShut {NoStop}%
\bibitem [{\citenamefont {Watson}\ \emph {et~al.}(2018)\citenamefont {Watson},
  \citenamefont {Philips}, \citenamefont {Kawakami}, \citenamefont {Ward},
  \citenamefont {Scarlino}, \citenamefont {Veldhorst}, \citenamefont {Savage},
  \citenamefont {Lagally}, \citenamefont {Friesen}, \citenamefont
  {Coppersmith},\ and\ \citenamefont {et~al.}}]{Watson_2018}%
  \BibitemOpen
  \bibfield  {author} {\bibinfo {author} {\bibfnamefont {T.~F.}\ \bibnamefont
  {Watson}}, \bibinfo {author} {\bibfnamefont {S.~G.~J.}\ \bibnamefont
  {Philips}}, \bibinfo {author} {\bibfnamefont {E.}~\bibnamefont {Kawakami}},
  \bibinfo {author} {\bibfnamefont {D.~R.}\ \bibnamefont {Ward}}, \bibinfo
  {author} {\bibfnamefont {P.}~\bibnamefont {Scarlino}}, \bibinfo {author}
  {\bibfnamefont {M.}~\bibnamefont {Veldhorst}}, \bibinfo {author}
  {\bibfnamefont {D.~E.}\ \bibnamefont {Savage}}, \bibinfo {author}
  {\bibfnamefont {M.~G.}\ \bibnamefont {Lagally}}, \bibinfo {author}
  {\bibfnamefont {M.}~\bibnamefont {Friesen}}, \bibinfo {author} {\bibfnamefont
  {S.~N.}\ \bibnamefont {Coppersmith}},\ and\ \bibinfo {author} {\bibnamefont
  {et~al.}},\ }\bibfield  {title} {\bibinfo {title} {A programmable two-qubit
  quantum processor in silicon},\ }\href {https://doi.org/10.1038/nature25766}
  {\bibfield  {journal} {\bibinfo  {journal} {Nature (London)}\ }\textbf
  {\bibinfo {volume} {555}},\ \bibinfo {pages} {633} (\bibinfo {year}
  {2018})}\BibitemShut {NoStop}%
\bibitem [{\citenamefont {Zajac}\ \emph {et~al.}(2017)\citenamefont {Zajac},
  \citenamefont {Sigillito}, \citenamefont {Russ}, \citenamefont {Borjans},
  \citenamefont {Taylor}, \citenamefont {Burkard},\ and\ \citenamefont
  {Petta}}]{Zajac_2017}%
  \BibitemOpen
  \bibfield  {author} {\bibinfo {author} {\bibfnamefont {D.~M.}\ \bibnamefont
  {Zajac}}, \bibinfo {author} {\bibfnamefont {A.~J.}\ \bibnamefont
  {Sigillito}}, \bibinfo {author} {\bibfnamefont {M.}~\bibnamefont {Russ}},
  \bibinfo {author} {\bibfnamefont {F.}~\bibnamefont {Borjans}}, \bibinfo
  {author} {\bibfnamefont {J.~M.}\ \bibnamefont {Taylor}}, \bibinfo {author}
  {\bibfnamefont {G.}~\bibnamefont {Burkard}},\ and\ \bibinfo {author}
  {\bibfnamefont {J.~R.}\ \bibnamefont {Petta}},\ }\bibfield  {title} {\bibinfo
  {title} {Resonantly driven cnot gate for electron spins},\ }\href
  {https://doi.org/10.1126/science.aao5965} {\bibfield  {journal} {\bibinfo
  {journal} {Science}\ }\textbf {\bibinfo {volume} {359}},\ \bibinfo {pages}
  {439} (\bibinfo {year} {2017})}\BibitemShut {NoStop}%
\bibitem [{\citenamefont {Brunner}\ \emph {et~al.}(2011)\citenamefont
  {Brunner}, \citenamefont {Shin}, \citenamefont {Obata}, \citenamefont
  {Pioro-Ladri\`ere}, \citenamefont {Kubo}, \citenamefont {Yoshida},
  \citenamefont {Taniyama}, \citenamefont {Tokura},\ and\ \citenamefont
  {Tarucha}}]{PhysRevLett.107.146801}%
  \BibitemOpen
  \bibfield  {author} {\bibinfo {author} {\bibfnamefont {R.}~\bibnamefont
  {Brunner}}, \bibinfo {author} {\bibfnamefont {Y.-S.}\ \bibnamefont {Shin}},
  \bibinfo {author} {\bibfnamefont {T.}~\bibnamefont {Obata}}, \bibinfo
  {author} {\bibfnamefont {M.}~\bibnamefont {Pioro-Ladri\`ere}}, \bibinfo
  {author} {\bibfnamefont {T.}~\bibnamefont {Kubo}}, \bibinfo {author}
  {\bibfnamefont {K.}~\bibnamefont {Yoshida}}, \bibinfo {author} {\bibfnamefont
  {T.}~\bibnamefont {Taniyama}}, \bibinfo {author} {\bibfnamefont
  {Y.}~\bibnamefont {Tokura}},\ and\ \bibinfo {author} {\bibfnamefont
  {S.}~\bibnamefont {Tarucha}},\ }\bibfield  {title} {\bibinfo {title}
  {Two-qubit gate of combined single-spin rotation and interdot spin exchange
  in a double quantum dot},\ }\href
  {https://doi.org/10.1103/PhysRevLett.107.146801} {\bibfield  {journal}
  {\bibinfo  {journal} {Phys. Rev. Lett.}\ }\textbf {\bibinfo {volume} {107}},\
  \bibinfo {pages} {146801} (\bibinfo {year} {2011})}\BibitemShut {NoStop}%
\bibitem [{\citenamefont {Martins}\ \emph {et~al.}(2016)\citenamefont
  {Martins}, \citenamefont {Malinowski}, \citenamefont {Nissen}, \citenamefont
  {Barnes}, \citenamefont {Fallahi}, \citenamefont {Gardner}, \citenamefont
  {Manfra}, \citenamefont {Marcus},\ and\ \citenamefont
  {Kuemmeth}}]{PhysRevLett.116.116801}%
  \BibitemOpen
  \bibfield  {author} {\bibinfo {author} {\bibfnamefont {F.}~\bibnamefont
  {Martins}}, \bibinfo {author} {\bibfnamefont {F.~K.}\ \bibnamefont
  {Malinowski}}, \bibinfo {author} {\bibfnamefont {P.~D.}\ \bibnamefont
  {Nissen}}, \bibinfo {author} {\bibfnamefont {E.}~\bibnamefont {Barnes}},
  \bibinfo {author} {\bibfnamefont {S.}~\bibnamefont {Fallahi}}, \bibinfo
  {author} {\bibfnamefont {G.~C.}\ \bibnamefont {Gardner}}, \bibinfo {author}
  {\bibfnamefont {M.~J.}\ \bibnamefont {Manfra}}, \bibinfo {author}
  {\bibfnamefont {C.~M.}\ \bibnamefont {Marcus}},\ and\ \bibinfo {author}
  {\bibfnamefont {F.}~\bibnamefont {Kuemmeth}},\ }\bibfield  {title} {\bibinfo
  {title} {Noise suppression using symmetric exchange gates in spin qubits},\
  }\href {https://doi.org/10.1103/PhysRevLett.116.116801} {\bibfield  {journal}
  {\bibinfo  {journal} {Phys. Rev. Lett.}\ }\textbf {\bibinfo {volume} {116}},\
  \bibinfo {pages} {116801} (\bibinfo {year} {2016})}\BibitemShut {NoStop}%
\bibitem [{\citenamefont {Reed}\ \emph {et~al.}(2016)\citenamefont {Reed},
  \citenamefont {Maune}, \citenamefont {Andrews}, \citenamefont {Borselli},
  \citenamefont {Eng}, \citenamefont {Jura}, \citenamefont {Kiselev},
  \citenamefont {Ladd}, \citenamefont {Merkel}, \citenamefont {Milosavljevic},
  \citenamefont {Pritchett}, \citenamefont {Rakher}, \citenamefont {Ross},
  \citenamefont {Schmitz}, \citenamefont {Smith}, \citenamefont {Wright},
  \citenamefont {Gyure},\ and\ \citenamefont
  {Hunter}}]{PhysRevLett.116.110402}%
  \BibitemOpen
  \bibfield  {author} {\bibinfo {author} {\bibfnamefont {M.~D.}\ \bibnamefont
  {Reed}}, \bibinfo {author} {\bibfnamefont {B.~M.}\ \bibnamefont {Maune}},
  \bibinfo {author} {\bibfnamefont {R.~W.}\ \bibnamefont {Andrews}}, \bibinfo
  {author} {\bibfnamefont {M.~G.}\ \bibnamefont {Borselli}}, \bibinfo {author}
  {\bibfnamefont {K.}~\bibnamefont {Eng}}, \bibinfo {author} {\bibfnamefont
  {M.~P.}\ \bibnamefont {Jura}}, \bibinfo {author} {\bibfnamefont {A.~A.}\
  \bibnamefont {Kiselev}}, \bibinfo {author} {\bibfnamefont {T.~D.}\
  \bibnamefont {Ladd}}, \bibinfo {author} {\bibfnamefont {S.~T.}\ \bibnamefont
  {Merkel}}, \bibinfo {author} {\bibfnamefont {I.}~\bibnamefont
  {Milosavljevic}}, \bibinfo {author} {\bibfnamefont {E.~J.}\ \bibnamefont
  {Pritchett}}, \bibinfo {author} {\bibfnamefont {M.~T.}\ \bibnamefont
  {Rakher}}, \bibinfo {author} {\bibfnamefont {R.~S.}\ \bibnamefont {Ross}},
  \bibinfo {author} {\bibfnamefont {A.~E.}\ \bibnamefont {Schmitz}}, \bibinfo
  {author} {\bibfnamefont {A.}~\bibnamefont {Smith}}, \bibinfo {author}
  {\bibfnamefont {J.~A.}\ \bibnamefont {Wright}}, \bibinfo {author}
  {\bibfnamefont {M.~F.}\ \bibnamefont {Gyure}},\ and\ \bibinfo {author}
  {\bibfnamefont {A.~T.}\ \bibnamefont {Hunter}},\ }\bibfield  {title}
  {\bibinfo {title} {Reduced sensitivity to charge noise in semiconductor spin
  qubits via symmetric operation},\ }\href
  {https://doi.org/10.1103/PhysRevLett.116.110402} {\bibfield  {journal}
  {\bibinfo  {journal} {Phys. Rev. Lett.}\ }\textbf {\bibinfo {volume} {116}},\
  \bibinfo {pages} {110402} (\bibinfo {year} {2016})}\BibitemShut {NoStop}%
\bibitem [{\citenamefont {Bertrand}\ \emph {et~al.}(2015)\citenamefont
  {Bertrand}, \citenamefont {Flentje}, \citenamefont {Takada}, \citenamefont
  {Yamamoto}, \citenamefont {Tarucha}, \citenamefont {Ludwig}, \citenamefont
  {Wieck}, \citenamefont {B\"auerle},\ and\ \citenamefont
  {Meunier}}]{PhysRevLett.115.096801}%
  \BibitemOpen
  \bibfield  {author} {\bibinfo {author} {\bibfnamefont {B.}~\bibnamefont
  {Bertrand}}, \bibinfo {author} {\bibfnamefont {H.}~\bibnamefont {Flentje}},
  \bibinfo {author} {\bibfnamefont {S.}~\bibnamefont {Takada}}, \bibinfo
  {author} {\bibfnamefont {M.}~\bibnamefont {Yamamoto}}, \bibinfo {author}
  {\bibfnamefont {S.}~\bibnamefont {Tarucha}}, \bibinfo {author} {\bibfnamefont
  {A.}~\bibnamefont {Ludwig}}, \bibinfo {author} {\bibfnamefont {A.~D.}\
  \bibnamefont {Wieck}}, \bibinfo {author} {\bibfnamefont {C.}~\bibnamefont
  {B\"auerle}},\ and\ \bibinfo {author} {\bibfnamefont {T.}~\bibnamefont
  {Meunier}},\ }\bibfield  {title} {\bibinfo {title} {Quantum manipulation of
  two-electron spin states in isolated double quantum dots},\ }\href
  {https://doi.org/10.1103/PhysRevLett.115.096801} {\bibfield  {journal}
  {\bibinfo  {journal} {Phys. Rev. Lett.}\ }\textbf {\bibinfo {volume} {115}},\
  \bibinfo {pages} {096801} (\bibinfo {year} {2015})}\BibitemShut {NoStop}%
\bibitem [{\citenamefont {Noiri}\ \emph {et~al.}(2021)\citenamefont {Noiri},
  \citenamefont {Takeda}, \citenamefont {Nakajima}, \citenamefont {Kobayashi},
  \citenamefont {Sammak}, \citenamefont {Scappucci},\ and\ \citenamefont
  {Tarucha}}]{noiri2021fast}%
  \BibitemOpen
  \bibfield  {author} {\bibinfo {author} {\bibfnamefont {A.}~\bibnamefont
  {Noiri}}, \bibinfo {author} {\bibfnamefont {K.}~\bibnamefont {Takeda}},
  \bibinfo {author} {\bibfnamefont {T.}~\bibnamefont {Nakajima}}, \bibinfo
  {author} {\bibfnamefont {T.}~\bibnamefont {Kobayashi}}, \bibinfo {author}
  {\bibfnamefont {A.}~\bibnamefont {Sammak}}, \bibinfo {author} {\bibfnamefont
  {G.}~\bibnamefont {Scappucci}},\ and\ \bibinfo {author} {\bibfnamefont
  {S.}~\bibnamefont {Tarucha}},\ }\href@noop {} {\bibinfo {title} {Fast
  universal quantum control above the fault-tolerance threshold in silicon}}
  (\bibinfo {year} {2021}),\ \Eprint {https://arxiv.org/abs/2108.02626}
  {arXiv:2108.02626 [quant-ph]} \BibitemShut {NoStop}%
\bibitem [{\citenamefont {Xue}\ \emph {et~al.}(2021)\citenamefont {Xue},
  \citenamefont {Russ}, \citenamefont {Samkharadze}, \citenamefont {Undseth},
  \citenamefont {Sammak}, \citenamefont {Scappucci},\ and\ \citenamefont
  {Vandersypen}}]{xue2021computing}%
  \BibitemOpen
  \bibfield  {author} {\bibinfo {author} {\bibfnamefont {X.}~\bibnamefont
  {Xue}}, \bibinfo {author} {\bibfnamefont {M.}~\bibnamefont {Russ}}, \bibinfo
  {author} {\bibfnamefont {N.}~\bibnamefont {Samkharadze}}, \bibinfo {author}
  {\bibfnamefont {B.}~\bibnamefont {Undseth}}, \bibinfo {author} {\bibfnamefont
  {A.}~\bibnamefont {Sammak}}, \bibinfo {author} {\bibfnamefont
  {G.}~\bibnamefont {Scappucci}},\ and\ \bibinfo {author} {\bibfnamefont
  {L.~M.~K.}\ \bibnamefont {Vandersypen}},\ }\href@noop {} {\bibinfo {title}
  {Computing with spin qubits at the surface code error threshold}} (\bibinfo
  {year} {2021}),\ \Eprint {https://arxiv.org/abs/2107.00628} {arXiv:2107.00628
  [quant-ph]} \BibitemShut {NoStop}%
\bibitem [{\citenamefont {Russ}\ \emph {et~al.}(2018)\citenamefont {Russ},
  \citenamefont {Zajac}, \citenamefont {Sigillito}, \citenamefont {Borjans},
  \citenamefont {Taylor}, \citenamefont {Petta},\ and\ \citenamefont
  {Burkard}}]{Russ_2018}%
  \BibitemOpen
  \bibfield  {author} {\bibinfo {author} {\bibfnamefont {M.}~\bibnamefont
  {Russ}}, \bibinfo {author} {\bibfnamefont {D.~M.}\ \bibnamefont {Zajac}},
  \bibinfo {author} {\bibfnamefont {A.~J.}\ \bibnamefont {Sigillito}}, \bibinfo
  {author} {\bibfnamefont {F.}~\bibnamefont {Borjans}}, \bibinfo {author}
  {\bibfnamefont {J.~M.}\ \bibnamefont {Taylor}}, \bibinfo {author}
  {\bibfnamefont {J.~R.}\ \bibnamefont {Petta}},\ and\ \bibinfo {author}
  {\bibfnamefont {G.}~\bibnamefont {Burkard}},\ }\bibfield  {title} {\bibinfo
  {title} {High-fidelity quantum gates in {S}i/{S}i{G}e double quantum dots},\
  }\href {http://dx.doi.org/10.1103/PhysRevB.97.085421} {\bibfield  {journal}
  {\bibinfo  {journal} {Phys. Rev. B}\ }\textbf {\bibinfo {volume} {97}},\
  \bibinfo {pages} {085421} (\bibinfo {year} {2018})}\BibitemShut {NoStop}%
\bibitem [{\citenamefont {Golovach}\ \emph {et~al.}(2006)\citenamefont
  {Golovach}, \citenamefont {Borhani},\ and\ \citenamefont
  {Loss}}]{PhysRevB.74.165319}%
  \BibitemOpen
  \bibfield  {author} {\bibinfo {author} {\bibfnamefont {V.~N.}\ \bibnamefont
  {Golovach}}, \bibinfo {author} {\bibfnamefont {M.}~\bibnamefont {Borhani}},\
  and\ \bibinfo {author} {\bibfnamefont {D.}~\bibnamefont {Loss}},\ }\bibfield
  {title} {\bibinfo {title} {Electric-dipole-induced spin resonance in quantum
  dots},\ }\href {https://doi.org/10.1103/PhysRevB.74.165319} {\bibfield
  {journal} {\bibinfo  {journal} {Phys. Rev. B}\ }\textbf {\bibinfo {volume}
  {74}},\ \bibinfo {pages} {165319} (\bibinfo {year} {2006})}\BibitemShut
  {NoStop}%
\bibitem [{\citenamefont {Tokura}\ \emph {et~al.}(2006)\citenamefont {Tokura},
  \citenamefont {van~der Wiel}, \citenamefont {Obata},\ and\ \citenamefont
  {Tarucha}}]{PhysRevLett.96.047202}%
  \BibitemOpen
  \bibfield  {author} {\bibinfo {author} {\bibfnamefont {Y.}~\bibnamefont
  {Tokura}}, \bibinfo {author} {\bibfnamefont {W.~G.}\ \bibnamefont {van~der
  Wiel}}, \bibinfo {author} {\bibfnamefont {T.}~\bibnamefont {Obata}},\ and\
  \bibinfo {author} {\bibfnamefont {S.}~\bibnamefont {Tarucha}},\ }\bibfield
  {title} {\bibinfo {title} {Coherent single electron spin control in a
  slanting zeeman field},\ }\href
  {https://doi.org/10.1103/PhysRevLett.96.047202} {\bibfield  {journal}
  {\bibinfo  {journal} {Phys. Rev. Lett.}\ }\textbf {\bibinfo {volume} {96}},\
  \bibinfo {pages} {047202} (\bibinfo {year} {2006})}\BibitemShut {NoStop}%
\bibitem [{\citenamefont {Pedersen}\ \emph {et~al.}(2007)\citenamefont
  {Pedersen}, \citenamefont {Møller},\ and\ \citenamefont
  {Mølmer}}]{Pedersen_2007}%
  \BibitemOpen
  \bibfield  {author} {\bibinfo {author} {\bibfnamefont {L.~H.}\ \bibnamefont
  {Pedersen}}, \bibinfo {author} {\bibfnamefont {N.~M.}\ \bibnamefont
  {Møller}},\ and\ \bibinfo {author} {\bibfnamefont {K.}~\bibnamefont
  {Mølmer}},\ }\bibfield  {title} {\bibinfo {title} {Fidelity of quantum
  operations},\ }\href {https://doi.org/10.1016/j.physleta.2007.02.069}
  {\bibfield  {journal} {\bibinfo  {journal} {Phys. Lett. A}\ }\textbf
  {\bibinfo {volume} {367}},\ \bibinfo {pages} {47} (\bibinfo {year}
  {2007})}\BibitemShut {NoStop}%
\bibitem [{\citenamefont {Mizuta}\ \emph {et~al.}(2017)\citenamefont {Mizuta},
  \citenamefont {Otxoa}, \citenamefont {Betz},\ and\ \citenamefont
  {Gonzalez-Zalba}}]{PhysRevB.95.045414}%
  \BibitemOpen
  \bibfield  {author} {\bibinfo {author} {\bibfnamefont {R.}~\bibnamefont
  {Mizuta}}, \bibinfo {author} {\bibfnamefont {R.~M.}\ \bibnamefont {Otxoa}},
  \bibinfo {author} {\bibfnamefont {A.~C.}\ \bibnamefont {Betz}},\ and\
  \bibinfo {author} {\bibfnamefont {M.~F.}\ \bibnamefont {Gonzalez-Zalba}},\
  }\bibfield  {title} {\bibinfo {title} {Quantum and tunneling capacitance in
  charge and spin qubits},\ }\href {https://doi.org/10.1103/PhysRevB.95.045414}
  {\bibfield  {journal} {\bibinfo  {journal} {Phys. Rev. B}\ }\textbf {\bibinfo
  {volume} {95}},\ \bibinfo {pages} {045414} (\bibinfo {year}
  {2017})}\BibitemShut {NoStop}%
\bibitem [{\citenamefont {Heinz}\ and\ \citenamefont
  {Burkard}(2021{\natexlab{a}})}]{PhysRevB.104.045420}%
  \BibitemOpen
  \bibfield  {author} {\bibinfo {author} {\bibfnamefont {I.}~\bibnamefont
  {Heinz}}\ and\ \bibinfo {author} {\bibfnamefont {G.}~\bibnamefont
  {Burkard}},\ }\bibfield  {title} {\bibinfo {title} {Crosstalk analysis for
  single-qubit and two-qubit gates in spin qubit arrays},\ }\href
  {https://doi.org/10.1103/PhysRevB.104.045420} {\bibfield  {journal} {\bibinfo
   {journal} {Phys. Rev. B}\ }\textbf {\bibinfo {volume} {104}},\ \bibinfo
  {pages} {045420} (\bibinfo {year} {2021}{\natexlab{a}})}\BibitemShut
  {NoStop}%
\bibitem [{\citenamefont {Heinz}\ and\ \citenamefont
  {Burkard}(2021{\natexlab{b}})}]{heinz2021crosstalk}%
  \BibitemOpen
  \bibfield  {author} {\bibinfo {author} {\bibfnamefont {I.}~\bibnamefont
  {Heinz}}\ and\ \bibinfo {author} {\bibfnamefont {G.}~\bibnamefont
  {Burkard}},\ }\href@noop {} {\bibinfo {title} {Crosstalk analysis for
  simultaneously driven two-qubit gates in spin qubit arrays}} (\bibinfo {year}
  {2021}{\natexlab{b}}),\ \Eprint {https://arxiv.org/abs/2111.10174}
  {arXiv:2111.10174 [cond-mat.mes-hall]} \BibitemShut {NoStop}%
\bibitem [{\citenamefont {Russ}\ \emph {et~al.}(2020)\citenamefont {Russ},
  \citenamefont {P{\'{e}}terfalvi},\ and\ \citenamefont {Burkard}}]{Russ_2020}%
  \BibitemOpen
  \bibfield  {author} {\bibinfo {author} {\bibfnamefont {M.}~\bibnamefont
  {Russ}}, \bibinfo {author} {\bibfnamefont {C.~G.}\ \bibnamefont
  {P{\'{e}}terfalvi}},\ and\ \bibinfo {author} {\bibfnamefont {G.}~\bibnamefont
  {Burkard}},\ }\bibfield  {title} {\bibinfo {title} {Theory of valley-resolved
  spectroscopy of a si triple quantum dot coupled to a microwave resonator},\
  }\href {https://doi.org/10.1088/1361-648x/ab613f} {\bibfield  {journal}
  {\bibinfo  {journal} {Journal of Physics: Condensed Matter}\ }\textbf
  {\bibinfo {volume} {32}},\ \bibinfo {pages} {165301} (\bibinfo {year}
  {2020})}\BibitemShut {NoStop}%
\end{thebibliography}%

\end{document}